\documentstyle[preprint,aps]{revtex}

\def\journal#1, #2, #3, 1#4#5#6{
    {\sl #1~}{\bf #2}, #3 (1#4#5#6)}
\def\pr{\journal Phys. Rev., }

\def\prd{\journal Phys. Rev. D, }
\def\prl{\journal Phys. Rev. Lett., }

\def\pl{\journal Phys. Lett., }

\def\beq{\begin{equation}}
\def\eeq{\end{equation}}
\def\CC{{\cal C}}
\def\DD{{\cal D}}
\def\ie{{\it i.e., }}
\def\bfr{\hbox{\bf r}}
\def\a{\alpha}
\def\eps{\epsilon}
\def\la{\lambda}
\begin{document}
\preprint{UdeM-GPP-TH-95-27}

\title{Altered Stats: Two anyons via path integrals for multiply connected
spaces}
\author{P. Girard\footnote{e-mail: pgirard@lps.umontreal.ca} and
R. MacKenzie\footnote{e-mail: rbmack@lps.umontreal.ca}}
\address{Laboratoire de Physique Nucl\'eaire, Universit\'e de Montr\'eal,
C.P. 6128, succ. centreville, Montr\'eal, Qu\'ebec, Canada, H3C 3J7}
\maketitle

\begin{abstract}
\widetext

We apply the formalism of path integrals in multiply connected spaces to
the problem of two anyons.

\narrowtext \end{abstract}
\bigskip\bigskip
%\centerline{hep-th/aabbccc}
\newpage
\baselineskip22pt
Quantum mechanical systems of identical particles provide an elegant
situation in which one can apply the ideas of path integrals in multiply
connected configuration spaces \cite{schul1,laimor}.
The reason is clear if one
permits oneself to ignore coincident points in the configuration space,
\ie configurations where two or more of the particles are at the same
position \cite{bousor}. In that case, non-contractible closed loops in
configuration  space can easily be constructed. For example, a continuous
evolution of the system involving an interchange of particles forms a
closed path due to the identical nature of the particles, yet the path is
non-contractible due to the interchange.

According to the general rules of path integrals in multiply connected
configuration space,\footnote{For an excellent discussion, see
Ref. \cite{schul3}.} one can divide the space of paths into homotopy
classes, and rewrite the path integral as a sum of
sub-integrals, each of which is a
path integral over one such class:
\beq
\int\DD x(t)\,e^{iS}=\sum_n\int_{{\CC_n}}\DD x(t)\,e^{iS},
\label{one}
\eeq
where $\CC_n$ denotes the homotopy class labelled by $n$. Since the main
requirement of the path integral is that it satisfy the Schroedinger
equation, and since each sub-integral in (\ref{one}) is itself a solution,
any linear combination of the sub-integrals
will still satisfy the Schroedinger equation. For consistency of the path
integral, we must impose one requirement on the coefficients: that they
form a commutative
representation of the first homotopy group of the configuration space.

A very instructive example of the above ideas, due to its simplicity, is
provided by a
free particle constrained to move on a circle \cite{schul1,schul3}. There,
the paths are classified according to the net number of windings around the
circle. The topology involved can be regarded from the point of view of the
simply-connected covering space of the configuration space, which is simply
a line. Each point on the circle is the image of an infinity of points on
the line, and different homotopy classes on the circle correspond to
classes of paths on the line
with different pre-images of the final point under the
mapping between the circle and line. The sub-integral corresponding to any
homotopy class on the circle, viewed in the covering space, is an ordinary
path integral between the initial point and the corresponding final point
on the line. The path integral for the circle, then, becomes an infinite
sum of free-particle path integrals in one dimension.

One can, as described above, add relative phases $\exp-in\alpha$, where $n$
denotes the number of times a path winds around the circle, to the
sub-integrals; such liberty, in fact, corresponds to threading a magnetic
flux $\Phi=\a/e$ through the circle \cite{schul1}.

Applying these ideas to systems of identical particles,
the above reasoning was used to demonstrate that only Bose and
Fermi statistics are possible in three space
dimensions \cite{laimor}. The reason is essentially that
the first homotopy group of the configuration space of $N$ identical
particles in 3 dimensions is the permutation group $S_N$, which has only
two commutative representations: the trivial representation and that
representation which
assigns $\pm1$ to a path according to whether it corresponds to an even or
odd permutation of the particles. The former describes Bose statistics, the
latter, Fermi statistics.

The work of Ref. \cite{laimor}\ can in fact be turned around slightly:
using a path integral formulation, a system of identical spinless fermions
could be described by bosons, by dividing the path integral into its
homotopy classes and adding a factor $-1$ to all paths corresponding to odd
permutations of the particles. Although this is true in principle, the
division of the path integral into homotopy classes appears to be rather
difficult, and to our knowledge such an approach has not been put into
practice.

More recently, it was realized by completely independent means that
fractional statistics (anyons) is allowed in two space dimensions
\cite{leimyr,wilczek1,wilczek2}.\footnote{For a comprehensive review
including many applications, see Ref. \cite{wilczek3}.}
A subsequent re-examination of the ideas of Ref. \cite{laimor}\
in two space dimensions \cite{wu}\ revealed in an elegant way that indeed
the possibility of fractional statistics might have been anticipated via
path integrals: the relevant homotopy group is a braid group (rather than a
permutation group), and the commutative representations of such groups
allow for any relative phase between different homotopy classes.

Anyons are normally described by adding fictitious (``statistical'')
charges and magnetic fluxes to ``normal'' particles (bosons or fermions) in
such a way that no new forces are introduced but so that windings of each
particle around the others are counted in an appropriate way. As in three
dimensions, a path integral approach provides, in principle, an alternative
means of altering the statistics. One can imagine dividing the path
integral for identical {\sl bosons} into homotopy classes, and adding
relative phases between different classes to change the way different
windings interfere with one another in a way appropriate for the desired
statistics.

Once again, putting this program into practice is extremely difficult. The
problem is that one needs a set of coordinates which retain a memory of the
relative windings of the particles. (For instance, Cartesian coordinates
are clearly inappropriate since the coordinates of a given configuration of
particles are unique; the winding of particles around one another is lost
in the Cartesian description of the motion.) For the case of two particles
the resolution is obvious: if, after disposing of trivial center-of-mass
motion, the relative motion is described in polar coordinates, the angular
coordinate measures the winding of the two particles around one another.
The path integral can thus be divided into homotopy sectors where the final
point in configuration space is described by final angles differing by a
multiple of $\pi$.

Indeed, a program along these lines has been put into practice in the
context of a path integral description of the Aharonov-Bohm effect
\cite{gersin,arovas},
which has obvious and well-known similarities with anyons
\cite{wilczek2}, and in the context of the statistical mechanics of anyons
\cite{myrola}. However, these discussions of the Aharonov-Bohm effect are
in a sense ``less topological'' than, for instance, the treatment outlined
above of the path integral for a particle on a circle \cite{schul1,schul3}.
Essentially, a (non-topological) path integral in polar
coordinates is written down
for a particle moving in the plane \cite{edwgul,peaino}, and
a topological constraint on the path integral is added after the fact. The
intermediate angular positions are integrated only over the interval
$[0,2\pi]$, in contrast with the general approach discussed in
Refs. \cite{schul1,schul3}.

In this paper, we present an alternative approach to the problem of the
relative motion of two anyons, which has a certain aesthetic advantage over
that described above, in that it is more faithful to the topology of the
situation. The approach consists in regarding the motion from the point of
view of the simply-connected covering space for the plane with the origin
excised, which can be visualized either as a sort of spiral staircase or as
a half plane, and performing the path integral on this space, allowing the
angle of intermediate steps in a path to take on any real values in a
manner exactly analogous to Refs. \cite{schul1,schul3}, rather than
restricting them to the range $[0,2\pi]$. Our approach is similar in spirit
to that of Khandekar, Bhagwat and Wiegel \cite{khabhawie}, who discussed
the problem of the motion of a particle in a plane with the origin removed
from an equally topological point of view,
although the details of our calculation differ from theirs.

We begin with the Lagrangian describing the relative motion of two bosons
of unit mass moving in the plane with the origin removed,
in polar coordinates:
\beq
L={1\over2}({\dot r}^2+r^2{\dot\theta}^2).
\label{two}
\eeq
Rather than adding an explicit interaction term to (\ref{two}) to alter
the statistics of the particles, we will
introduce relative phases between the different homotopy classes of the
path integral, as described above.

The propagator for motion from a point $\bfr'=(r',\theta')$ to
$\bfr''=(r'',\theta'')$ in a time interval $T$ is the sum over all paths
from the initial to the final point:
\beq
K(\bfr'';\bfr';T)=\int_{\bfr'}^{\bfr''}\DD\bfr(t)\, e^{iS}.
\eeq
If we write the paths in polar
coordinates, $(r(t),\theta(t))$, then $(r(0),\theta(0))=(r',\theta')$, and
$(r(T),\theta(T))=(r'',\theta''+n\pi)\equiv\bfr''_n$,
reflecting the fact that changing
the final angle by an integer multiple of
$\pi$ describes the same final state: the paths between
a given initial and final configuration can be divided into homotopy
classes labelled by $n$.

We can therefore rewrite the path integral in the following way:
\beq
K(\bfr'';\bfr';T)=\sum_n\int_{\bfr'}^{\bfr''_n}\DD\bfr(t)\,
e^{iS}\equiv\sum_n K_n.
\label{four}
\eeq
We are now, according to Ref. \cite{schul1}, free to add phases between the
sub-integrals in (\ref{four}):
\beq
K\to K^{(\a)}=\sum_n e^{-in\a}K_n.
\label{five}
\eeq
The phases change the interference between different homotopy sectors,
reflecting altered statistics: if $\a=\pi$ (or any odd multiple of $\pi$),
the amplitudes of paths whose $n$'s differ by one subtract rather than
adding, as appropriate for Fermi statistics, while if $\a$ is any even
multiple of $\pi$, the additional phases in (\ref{five}) have no effect,
and the particles are bosons. For arbitrary $\a$, the particles are anyons.

We see that anyons can be described by perfectly conventional path
integrals within each homotopy sector (\ie without introducing the
statistical interaction), if the total propagator is computed with
appropriate phases between the different sectors.

Let us compute, then, a sub-integral, $K_n$. The computation is performed
in polar coordinates, which introduces certain subtleties. In particular,
it was shown in Ref. \cite{edwgul} that the most naive guess for the path
integral must be modified, due to a path integral version of operator
ordering ambiguities. In Ref. \cite{mclsch}, it was shown that a path
integral in curved space requires the inclusion of an additional
``effective potential'' term to the Hamiltonian; applying that result to
the case at hand (where polar coordinates introduce a nontrivial metric, so
the formalism of Ref. \cite{mclsch} applies directly), this effective
potential is $-1/(8r^2)$, and the
sub-integral becomes
\beq
K_n={1\over(2\pi i\eps)^N}\int\prod_{j=1}^{N-1}(r_j dr_j d\theta_j)
\exp i\sum_{j=1}^N\left({(r_j-r_{j-1})^2+({\bar r}_j)^2
(\theta_j-\theta_{j-1})^2\over2\eps}
+{\eps\over8({\bar r}_j)^2}\right),
\eeq
where the limit $N\to\infty$ will be understood throughout,
$\theta_0=\theta'$, $\theta_N=\theta''+n\pi$, and
${\bar r}_j$ is the average position for the interval $(j-1,j)$, which we
take to be the geometric mean $\sqrt{r_{j-1}r_j}$.

We can simplify the angular integrals considerably by the following trick.
First, we insert a factor $1=\int
d\theta_N\delta(\theta_N-(\theta''+n\pi))$. Second, we rewrite the
argument of the $\delta$-function in terms of the angular differences
$\Delta\theta_j\equiv\theta_j-\theta_{j-1}$; it becomes
$\Delta\theta_1+\dots+\Delta\theta_N-\delta\theta_n$, where
$\delta\theta_n\equiv\theta''+n\pi-\theta'$ is the total angular change
for the propagator for the $n$th sector. Third, we rewrite the Dirac
$\delta$-function as an exponential:
\beq
\delta\left(\sum_1^N\Delta\theta_j-\delta\theta_n\right)
=\int{d\la\over2\pi}
\exp i\la\left(\sum_1^N\Delta\theta_j-\delta\theta_n\right).
\eeq
The $N$ angular integrals can now be
rewritten in terms of integrals over the $N$ angular differences
$\Delta\theta_j$; the integrals so obtained are independant Gaussians and
we obtain
\beq
K_n={1\over(2\pi i\eps)^{N/2}}\int{d\la\over2\pi}e^{-i\la\delta\theta_n}
\int\prod_1^{N-1}(r_j dr_j)\prod_1^N({\bar r}_j)^{-2}
\exp{i\over2}\sum_1^N\left(
{(r_j-r_{j-1})^2\over\eps}-{(\la^2-1/4)\eps\over({\bar r}_j)^2}
\right).
\eeq
The integrand can be usefully rewritten in terms of modified Bessel
functions, using the asymptotic form for these; one finds
\beq
K_n={e^{i(r'^2+r''^2)/2\eps}\over(i\eps)^N}
\int{d\la\over2\pi}e^{-i\la\delta\theta_n}
\int\prod_1^{N-1}(r_j dr_j)
\prod_1^N I_{|\la|}\left(({\bar r}_j)^2/i\eps\right)
\exp i\sum_1^{N-1}{r_j}^2/\eps.
\eeq
The radial integrals can now be performed \cite{peaino,arovas}, yielding
(taking the limit $N\to\infty$)
\beq
K_n={e^{i(r'^2+r''^2)/2T}\over iT}
\int{d\la\over2\pi}e^{-i\la\delta\theta_n}
I_{|\la|}(-ir'r''/T),
\label{junk}
\eeq
our final result for the sub-integral.

Although we have derived this result for a system of two identical bosons,
it applies equally well to the problem of a single particle moving in a
punctured plane (which is also, of course, described by the Lagrangian
(\ref{two}), if we write $\delta\theta_n=\theta''+2n\pi-\theta'$, since for
a single particle we have periodicity of $2\pi$ rather than $\pi$. With
this change, our result (\ref{junk}) is then in complete agreement
with that of Ref. \cite{khabhawie}.

A simple check of this result is to sum the sub-integrals
(taking $\delta\theta_n=\theta''+2n\pi-\theta'$)
for all the
topological sectors; one must obtain the propagator for a free particle.
Instead, we will perform this sum with relative phases added in, thus
obtaining the propagator for a charged particle in the presence of a
flux tube; the free particle is obtained in the limit where the flux is
zero. The propagator (for $e\phi=\a$) is
\beq
K^{(\a)}=\sum_n e^{-in\a}K_n=
{e^{i(r'^2+r''^2)/2T}\over iT}
\int{d\la\over2\pi}\sum_n e^{-i(\la+\a/2\pi)2\pi n}
e^{-i\la(\theta''-\theta')}
I_{|\la|}(-ir'r''/T).
\label{ten}
\eeq
Shifting $\la\to\la-\a/2\pi$ and using the Poisson summation formula
$\sum_n e^{-i\la 2\pi n} =\sum_m \a(\la+m)$,
\beq
K^{(\a)}={e^{i(r'^2+r''^2)/2T}\over 2\pi iT}
\sum_m e^{i(m+\a/2\pi)(\theta''-\theta')}
I_{|m+\a/2\pi|}(-ir'r''/T).
\label{eleven}
\eeq

This is our final result for the propagator in the presence of a flux tube.
It differs from the result of, e.g., Ref. \cite{arovas}\footnote{Note that
Ref. \cite{arovas} uses a definition of $\a$ which differs by a factor
$2\pi$ from ours.} by an overall phase $\exp i(\a/2\pi)(\theta''-\theta')$.
This phase is in fact simply a choice of gauge for the problem
\cite{gersin}, and is related to whether or not one chooses to describe the
particle by a single-valued wave function (as in \cite{arovas}) or not (as
in this work). Alternatively, our method of describing the interaction of
the particle with the flux tube takes into account the additional phase for
paths which wind around the solenoid different numbers of times, but it
does not take into account the phase due to the change in angle between the
initial and final points; if we wanted, we could have taken this into
account by adding an overall phase $\exp -i\a(\theta''-\theta')/2\pi$ to
the propagator, in which case (\ref{eleven}) would be in complete agreement
with Ref. \cite{arovas}.

We can take the limit $\a\to0$ in (\ref{eleven}); the sum is easily
evaluated in terms of the generating function for modified Bessel functions
and one finds that the free propagator is indeed obtained.

For the application of these ideas to the problem of the relative motion of
two anyons, the change is trivial: different homotopy sectors now differ in
their final angle by multiples of $\pi$ (rather than $2\pi$), and the change
in angle for the $n$th sector is now $\theta_n=\theta''+n\pi-\theta'$.
The propagator becomes
\beq
K^{(\a)}={e^{i(r'^2+r''^2)/2T}\over \pi iT}
\sum_m e^{i(2m+\a/\pi)(\theta''-\theta')}
I_{|2m+\a/\pi|}(-ir'r''/T).
\eeq
As expected, one can easily verify that when $\a=0$ or
$\pi$ (corresponding to
bosons or fermions) the propagator is
\beq
K={1\over2\pi iT}\left(e^{i|\bfr''-\bfr'|^2/2T}\pm
e^{i|\bfr''+\bfr'|^2/2T}\right).
\eeq

Unfortunately, applying these ideas to systems whose configuration space is
more complicated ({\it e.g.,} three or more anyons), while in principle
straightforward, appears exceedingly difficult because of the
difficulty in finding a suitable set of coordinates which retains a memory
of the windings of particles around one another. While such coordinates
have been found for the case of three anyons, and in fact an expression
analogous to (\ref{ten}) can be found, in those coordinates the action is
sufficiently horrible that the path integral does not seem feasible.

In summary, we have given a derivation of the propagator for a particle in
the plane with origin removed which is in agreement with previous work, but
which is more faithful to the topology of the configuration space. We have
added permitted phases to different sectors of the path integral in order
to describe the Aharonov-Bohm effect and a system of two anyons.

\bigskip\bigskip
This work was supported by the Natural Science and Engineering Research Council
of Canada.

\end{document}